\documentclass[letterpaper]{article} 
\usepackage{aaai24}  
\usepackage{times}  
\usepackage{helvet}  
\usepackage{courier}  
\usepackage[hyphens]{url}  
\usepackage{graphicx} 
\usepackage{natbib}  
\usepackage{caption} 
\usepackage{xspace}
\usepackage{paralist}
\usepackage{nth}
\usepackage{subcaption}
\usepackage{booktabs}
\usepackage{xcolor}

\usepackage[draft,inline,nomargin,index]{fixme}
\fxsetup{theme=colorsig,mode=multiuser,inlineface=\itshape,envface=\itshape}
\FXRegisterAuthor{cw}{acw}{Christo}
\FXRegisterAuthor{jg}{ajg}{Jeffrey}
\FXRegisterAuthor{az}{aaz}{Aziz}
\FXRegisterAuthor{dh}{adh}{Desheng}
\FXRegisterAuthor{rr}{arr}{Ronald}
\FXRegisterAuthor{ng}{ang}{Nik}

\newcommand{\fulltitle}{Market or Markets? Investigating Google Search's Market Shares Under Horizontal and Vertical Segmentation}

\nocopyright



\newcommand{\eg}{e.g.,\ }
\newcommand{\etal}{et al.\xspace}
\newcommand{\ie}{i.e.,\ }

\newcommand{\aka}{a.k.a.\ }

\newcommand{\verticalsvolume}{21\xspace}
\newcommand{\verticalssession}{24\xspace}

\title{\fulltitle\footnote{This manuscript is an extended version of a paper published in the Proceedings of the International AAAI Conference on Web and Social Media (ICWSM) 2024~\cite{hu-2024-icwsm}. The original paper is available here: https://doi.org/10.1609/icwsm.v18i1.31340.}}

\author{
    Desheng Hu,\equalcontrib\textsuperscript{\rm 1}
    Jeffrey Gleason,\equalcontrib\textsuperscript{\rm 2}
    Muhammad Abu Bakar Aziz,\equalcontrib\textsuperscript{\rm 2}
    Alice Koeninger,\textsuperscript{\rm 2}\\
    Nikolas Guggenberger,\textsuperscript{\rm 3}
    Ronald E. Robertson,\textsuperscript{\rm 2, \rm 4}
    Christo Wilson\textsuperscript{\rm 2}
}
\affiliations{
    \textsuperscript{\rm 1}University of Zurich, Zurich, Switzerland\\
    \textsuperscript{\rm 2}Northeastern University, Boston MA, USA\\
    \textsuperscript{\rm 3}University of Houston, Houston TX, USA\\
    \textsuperscript{\rm 4}Stanford University, Stanford CA, USA\\
    desheng@ifi.uzh.ch, gleason.je@northeastern.edu, aziz.muh@northeastern.edu,\\
    nguggenb@central.uh,edu, rer@acm.org, cbw@ccs.neu.edu
}

\begin{document}

\maketitle

\begin{abstract}
  Is Google Search a monopoly with gatekeeping power? Regulators from the US, UK, and Europe have argued that it is based on the assumption that Google Search dominates the market for horizontal (\aka ``general'') web search. Google disputes this, claiming that competition extends to all vertical (\aka ``specialized'') search engines, and that under this market definition it does not have monopoly power.
  
  In this study we present the first analysis of Google Search's market share under 
  both horizontal and 
  vertical segmentation of online search. We leverage observational trace data collected from a panel of US residents that includes their web browsing history and copies of the Google Search Engine Result Pages they were shown. We observe that 
  Google Search receives 71.8\% of participants' queries when compared to other horizontal search engines, and that 
  participants' search sessions begin at Google greater than 50\% of the time in \verticalssession out of 30 vertical market segments (which comprise almost all of our participants' searches). Our results inform the consequential and ongoing debates about the market power of Google Search and the conceptualization of online markets in general.
\end{abstract}

\section{Introduction}
\label{sec:introduction}

Google is one of largest corporations in the world. In 2022, it reported \$282.8B in revenue and a 26\% profit margin~\cite{sec2022}. Its products are ubiquitous---for example, it owns the world's most popular video streaming service~\cite{statista-video-usa}, web browser~\cite{statcounter-browser-global}, online display advertising platform~\cite{haggin-2019-google,us_texas,srinivasan-2019-stlr}, navigation and mapping application~\cite{statista-maps-usa}, and smartphone operating system~\cite{statcounter-mobile-global}. Google operates in numerous additional markets, including email and cloud computing~\cite{house2020}.

Among Google's many products, its original product---Google Search---continues to be its most lucrative. In 2022, 57.4\% of Google's revenue (\$162 billion) came from ad sales in Google Search~\cite{sec2022}. Google Search's market share is estimated to be 61--80\% of desktop web searches in the US, and has been stable for over 14 years~\cite{statista-search-usa,statcounter-search-usa}.
The durability of Google Search's position in the market is due, at least in part, to its position as the default web search engine on the vast majority of smartphones, tablets (\eg it pays Apple \$12B per year to be the default on iOS devices \cite{house2020}), and web browsers (\eg it pays Mozilla an estimated \$450M per year to be the default in Firefox~\cite{lyons-2020-mozilla}). Indeed, the word ``google'' is synonymous with the act of searching the web~\cite{google-verb}.

Because of its conduct towards competitors, Google Search has been the focus of numerous antitrust inquiries and litigation over the last decade. Currently, the US Department of Justice (DoJ) and 46 US states are suing Google for unlawfully maintaining a monopoly over the \textit{horizontal search} market~\cite{us_doj,us_colorado}---defined as the market for ``general search services'' that index the public web and return results for any query. The DoJ and the attorneys general allege that Google uses ``exclusionary default agreements'' with third parties like browser vendors, mobile device manufacturers, and cellular service providers to help maintain its monopoly. Other regulators and legislators have made the same allegations in the past~\cite{ftc2013,house2020,cma2020}, and collectively they
argue that Google's conduct harms consumers---especially when Google preferences its own products in search results. Indeed, in 2017, the European Commission (EC) found that Google abused its dominance in horizontal search by favoring its own comparison shopping service over those of competitors~\cite{ec_shopping_full}.

A critical facet of antitrust jurisprudence is the definition of the market for a good or service~\cite{market-definition}. Regulators and legislators have argued that Google Search dominates the market for horizontal search, within which, they claim, it competes with products like Microsoft Bing and DuckDuckGo~\cite{us_doj,us_colorado,cma2020,ftc2013,ec_shopping_full,house2020}. Google, however, disputes this market definition, claiming that competition extends to all horizontal and \textit{vertical search engines}---defined as search engines that specialize in one particular category of information or data from one particular service~\cite{ftc2013,house2020}. Google has argued that people search for ``news on Twitter, flights on Kayak and Expedia, restaurants on OpenTable, recommendations on Instagram and Pinterest'', and products on Amazon~\cite{google_doj_response}.
Similarly, in documents produced for the US House Subcommittee investigation, Google argued that estimates of its share of online search ``do not capture the full extent of Google's competition in search''~\cite{house2020}. To date, however, Google has provided no convincing evidence to back up its claim that it faces significant competition from vertical search engines.

In this study we present the first analysis of Google Search's market share under 
horizontal and 
vertical segmentation of online search. We leverage ecologically-valid, observational trace data collected from a panel of US residents over a five month period in 2020 that includes their web browsing history and complete copies of the Google Search Engine Result Pages (SERPs) they were shown.
To quantify horizontal market share, we compare participants' search behavior on Google Search, Bing, and other horizontal search engines. 
To quantify vertical market share, we identify searches carried out by participants on \textit{all} websites within our corpus, group all searches (on Google, on Bing, and on all other vertical search engines) into 90 vertical segments (\eg Shopping, Health and Wellness, News and Media), and compare participants' search behavior on and off Google products\footnote{``Google products'' includes participants' searches on Google's vertical search engines, \eg GMail and YouTube.} within each vertical segment. We also examine participants' propensity to switch between competing Google products, Microsoft products, and independent vertical competitors.

We observe that Google Search captures the most market share regardless of how the market for online search is segmented. Considering competition among horizontal search engines, 79.3\% of our participants used Google Search the majority of the time, and Google Search accounted for 71.8\% of all searches, which falls within publicly available estimates of Google Search's US desktop market share~\cite{statista-search-usa,statcounter-search-usa,comscore-search-usa-desktop,similarweb-search-usa-desktop}. Bing, in contrast, was preferred by 11.2\% of our participants and accounted for 23.8\% of all searches.

As it relates to competition between Google and vertical search engines, we find that Google's products receive over 50\% of participants' searches across \verticalsvolume of the top 30 market segments (which, collectively, account for 94.1\% of all searches performed by our participants). We also find that Google holds significant power as a \textit{gatekeeper} to independent vertical search engines~\cite{cma2020,house2020}. In \verticalssession of the top 30 market segments, participants began their search activity on a Google product more than 50\% of the time, sometimes followed by additional searches on an independent vertical search engine. Contrary to Google's assertions, our data suggests that participants do not treat Google Search and independent vertical search engines as substitutable.\footnote{``Substitutable'' products are equivalent to consumers. A Toyota sedan is substitutable for a Honda sedan, but a truck is not.} Further, our results highlight Google's power to steer users towards their own vertical search engines~\cite{jeffries-2020-google,gleason2023google}.

In summary, our work presents novel methods and analyses that inform consequential, ongoing debates about the market power of Google Search in particular, and the conceptualization of online markets in general. Our results speak to the prospects of ongoing antitrust litigation and the need for regulators to consider structural remedies---\eg separating Google Search from Google's vertical search engines, Android, and Chrome---and behavioral remedies---\eg prohibiting Google from signing exclusionary contracts with third-parties---to curtail the power of online intermediaries~\cite{khan-2019-clr,heidhues-2021-regulation}.

\section{Definitions}
\label{sec:definitions}

In its lawsuit against Google, the DoJ \etal define horizontal search engines as ```one-stop shops' consumers can use to search the internet for answers to a wide range of queries''~\cite{us_doj}. Others have used similar language to define horizontal search engines~\cite{ftc2013,ec_shopping_full,cma2020,house2020}.




In contrast, the DoJ \etal explain in their lawsuit that vertical search engines ``are not `one-stop shops' and cannot respond to all types of consumer queries, particularly navigational queries''~\cite{us_doj}. The staff at the US Federal Trade Commission (FTC) defined vertical search engines similarly, in 2012, as ``search engines focus[ed] on more narrowly-defined categories of content, such as product words''~\cite{ftc2013}. The US House Subcommittee concurred with these definition~\cite{house2020}.

We adopt this broad definition of vertical search and consider any website that supports search functionality, but is not a horizontal search engine, to be a vertical search engine. We chose this inclusive conceptualization of vertical search because it comports with Google's own assertions that it competes with all manners of websites (\eg social media, retailers, travel agencies, etc.) that support search functionality~\cite{google_doj_response,cma2020_appendixP,us_doj_defense}. We describe how we identified vertical search engines and quantified their usage in \S\,\ref{sec:vertical-search}.



In this study we examine Google and Bing's market share under vertical segmentation of the market for search, as well as their ability to function as gatekeepers to independent vertical search engines. The UK's Competition \& Markets Authority (CMA) defined ``gatekeepers'' as online platforms that ``mediat[e] relationships between consumers and businesses in a wide variety of markets''~\cite{cma2020}. The European Union's Digital Markets Act designates an online platform as a gatekeeper if ``(a) it has a significant impact on the internal market; (b) it provides a core platform service which is an important gateway for business users to reach end users; and (c) it enjoys an entrenched and durable position''~\cite{eu-dma}. Bipartisan antitrust legislation that includes a similar definition of ``covered platforms'' has been proposed in the US~\cite{aica-2021}.

In their lawsuit, the DoJ argues that Google Search is a gatekeeper to third-party websites in general, and vertical search engines in particular, due to its market share and its ability to answer navigational queries~\cite{us_doj}. Navigational queries describe searches for a specific website, often by name, followed by a click on the result link that points to this website~\cite{broder2002taxonomy,jansen2008determining}. Vertical search engines cannot answer navigational queries because they do not index the entirety of the web. We describe our approach for measuring gatekeeping power in \S\,\ref{sec:search-sessions}.

\section{Background}
\label{sec:background}

We now introduce related work that has studied online search engines and how our study draws from this literature.

\subsection{Quantifying Search Behavior}

There is a long history of scholars studying peoples' behavior on search engines. Early studies made foundational contributions to our understanding of how people (re)formulate queries and interact with Search Engine Result Pages (SERPs) using query logs from engines like AltaVista, Yahoo, and AOL~\cite{silverstein-1999-sigir,huang-2009-cikm,teevan-2007-sigir}. Unfortunately, these studies were limited to studying behavior on single search engines in isolation. Further, studies of query logs are rare today because search engines stopped sharing them in the wake of the AOL query log deanonymization debacle~\cite{barbaro-2006-aol}. That said, eye-tracking approaches continue to refine our understanding of how people interact with search engines~\cite{papoutsaki2017gazer}.

\subsubsection*{Navigational Search}

One influential study that emerged from the early search engine literature was a taxonomy of web search that included navigational, informational, and transactional queries~\cite{broder2002taxonomy}. Multiple approaches have been proposed to identify navigational queries. Jansen \etal identify navigational searches using a rules-based approach that checks whether a query contains
\begin{inparaenum}[(1)]
    \item a domain suffix or
    \item a company/organization name~\cite{jansen2008determining}.
\end{inparaenum}
In contrast, Teevan \etal identify navigational searches using a query's click entropy~\cite{teevan2011understanding}.
Previous studies have labeled 10--21\% of queries as navigational~\cite{jansen2008determining, teevan2011understanding}.

Navigational queries are a key facet of regulators' concerns about horizontal search engines. We adopt the Jansen \etal approach to identify navigational queries in this study~\cite{jansen2008determining} (see \S\,\ref{sec:assign-google}).

\subsubsection*{Search Sessions}
Jansen \etal define a search session as a ``series of interactions by the user toward addressing a single information need''~\cite{jansen2007defining}. The authors propose two approaches for identifying sessions:
\begin{inparaenum}[(1)]
    \item 30 minutes without a search, and
    \item query reformulation patterns. 
\end{inparaenum} The temporal approach produces a smaller number of sessions with a longer average length than the query reformulation approach. Many subsequent studies that model and analyze search behavior have used a 30 minute temporal cutoff to define a session boundary~\cite{downey2007models, downey2008understanding, white2009characterizing,hassan2014struggling}. One relevant finding from these studies is that only 4\% of sessions involve switching between horizontal search engines~\cite{white2009characterizing}. We adopt these methods to analyze our participants' search sessions.

\subsection{Auditing Search Engines}

\textit{Algorithm audits}~\cite{diakopoulos-2014-reporting,sandvig-2014-auditing} have utilized experimental and observational methods to collect and analyze data from search engines. Algorithm audits are motivated by the need to understand the impact of search engines' algorithmic choices---such as personalization~\cite{hannak-2013-filterbubbles,silver-2015-imc,robertson-www-2018}---on individuals and society. 
Audits have examined horizontal search engines like Google Search~\cite{diakopoulos-2018-vote,robertson2018auditing, hu2019auditing,robertson-2023-nature,scull2020dr,haddow2021inaccuracies} and vertical search engines like Amazon~\cite{juneja-2021-chi}, TaskRabbit~\cite{hannak-2017-cscw}, Indeed~\cite{chen-chi-2018}, Twitter~\cite{kulshrestha-2017-cscw}, and YouTube~\cite{hosseinmardi2020evaluating}.

\subsection{Competition in Search}

Several studies have focused on competition issues in the design of Google SERPs. Edelman and Lai exploited a natural experiment in which idiosyncratic differences in user queries determined whether Google displayed its Flight service on the SERP~\cite{edelman2016design}. They found that Google Flights increased paid click volume to travel agencies (\eg Expedia) by 65\% and decreased organic click volume by 55\%. Kim and Luca designed a controlled experiment to evaluate Google's decision to only include reviews from its own platform in the Local ``Onebox'' on the SERP~\cite{kim2019product}. They found that users preferred a Onebox that included reviews from competitors (\eg Yelp). Gleason \etal leveraged an observational dataset to find similar pairs of queries that triggered different SERP components~\cite{gleason2023google}. They found that Google's local, shopping, and image components decreased organic click-through rate (CTR) to third-party websites and that local and image components increased organic CTR to Google's own services.  

Our study relies on web browser extension-based data collection techniques that have been successfully used in many prior studies of horizontal search engines~\cite{robertson-www-2018,robertson2018auditing,robertson-2023-nature}.

\subsection{Domain Classification}

Automatically classifying websites and domains into topics or categories is a long-standing challenge. Numerous studies have proposed algorithms~\cite{zhang2004web,kwon2003text,sun2014web,buber2019web,lopez2017cbr} and features~\cite{mladenic1998turning,qi2009web,golub2005importance,shih2004using,utard2006link,camastra2015machine,lopez2019visual} for this task. Given the large number of approaches and dataset that are available for this task, recent studies have focused on comparing the relative accuracy of different classification approaches.~\cite{bruni2020website,hodvzic2016comparison,do2021phishing}. In this work, we adopt Fortiguard's domain to category mapping, based on the comprehensive evaluation in \citet{vallina2020mis} (see \S\,\ref{sec:mapping-websites}).

\section{Data and Methods}
\label{sec:methods}

In this section we present the datasets and methods that we used in our study. First, in \S\,\ref{sec:methods:participants}, we introduce the participant data we use throughout this study. 
Second, in \S\,\ref{sec:horizontal-search}, we discuss how we quantify search queries on horizontal search engines. 
Next, in \S\,\ref{sec:vertical-search}, we present our methodologies for identifying search queries on independent vertical search engines and grouping search queries into vertical market segments.
Finally, in \S\,\ref{sec:search-sessions}, we discuss our approach for clustering individual search queries into search sessions.

\subsection{Participant Data}
\label{sec:methods:participants}

Beginning in August 2020, we engaged the survey company YouGov to recruit a panel of US residents to take a survey and optionally install a browser extension we developed for Chrome and Firefox.\footnote{This study was IRB approved, see \S\,\ref{sec:broader} for details.} YouGov reached out to a nationally-representative sample of 2,000 people, of which $N=$ 926 completed the survey and installed the browser extension. We collected data from these participants' web browsers from August through December 2020. We adjusted all data collected from participants to be representative of the US adult population based on weights provided by YouGov.\footnote{\url{https://yougovplatform.zendesk.com/hc/en-gb/articles/360002975617-How-is-the-data-weighted}} Specifically, we multiplied counts of participants' online activities by their assigned weight.\footnote{Underrepresented and overrepresented participants are assigned higher and lower weights, respectively.} 
Additionally, based on self-reports, participants who installed the extension were slightly more likely to have high trust in Google Search and use it daily; we revisit this discrepancy in \S\,\ref{sec:limitations}.


To measure participants' web search behaviors on and off Google Search, our browser extension collected two types of passive, observational data from their web browsers: browsing history and \textit{snapshots} of Google SERPs. The browsing history data contains a record of every URL that participants loaded in their browser during our observation window and the timestamp at which each page load occurred. On average our participants loaded 296.6 URLs per day per participant ($\textrm{SD}$ = 49.2). The snapshot data contains the complete HTML of the SERPs that Google Search presented to participants in response to their queries. We collected and parsed 271,062 SERPs in total. On average our participants made 11.6 Google searches per day per participant ($\textrm{SD}$ = 1.7).


We observe that 97\% of the searches conducted by our participants on horizontal search engines occurred on Google and Bing (Yahoo and DuckDuckGo were the next two most frequently used). Thus, in the remainder of this study, we exclude activity that occurred on non-Google and non-Bing horizontal search engines.

\subsubsection{Parsing SERPs}

We examine Google SERPs broken down into vertical segments. To facilitate this segmentation (discussed below), we made use of the links that appeared in SERPs and participants' clicks on those links. We used the open source \texttt{WebSearcher} package to extract links from SERPs~\cite{robertson-2020-cj,robertson-websearcher}. 
On average we parsed 16.5 URLs per Google SERP ($\textrm{SD}$ = 9.9), which agrees with prior studies~\cite{robertson-www-2018,robertson2018auditing}.

\subsubsection{Click Measurement}

To identify which, if any, of the links in SERPs were clicked by participants, we examined the URLs that participants loaded immediately after performing a Google Search. Similar to prior work, our high-level approach to click measurement is to compare the exact URLs in a participant's browsing history within (1) thirty seconds and (2) three sequential URLs after performing a search to the exact set of URLs extracted from the SERP~\cite{flaxman-2016-poq,allen-2020-sciadv,guess-2020-nhb,guess2020sources}.

However, this exact comparison misses clicks on ads and URLs that redirect to a different URL. To address this, we identify ad clicks using the \texttt{gclid} URL parameter, which Google uses for conversion tracking and attribution.\footnote{https://support.google.com/google-ads/answer/9744275} Further, we identify clicks on redirected URLs by comparing the domains (\ie not the full URL) in a participant's browsing history within (1) thirty seconds and (2) three sequential URLs to the set of domains extracted from the SERP. To reduce false positives from this approach, we ignore any matches where the domain was included in the three URL visits prior to the search. This exclusion captures instances where a participant was browsing a website immediately before and after a Google search.

Using this approach we identified 103,599 clicks on SERPs and a per-SERP CTR of 38.2\% (similar to an estimate of 35\% from a recent industry report~\cite{seoclarity}). Our approach only detects the first click that participants made on SERPs---a limitation that we discuss in \S\,\ref{sec:limitations}.

\begin{table}[t]
  \centering
  \footnotesize
  
\begin{tabular}{lr} 
\toprule 
\textbf{Search Engine} & \textbf{Weighted Sum} \\ 
\midrule
Google & 364353 \\
Bing   & 120838 \\  
Yahoo  & 10138  \\ 
DuckDuckGo & 9772\\
Ecosia & 2282 \\
Info & 261 \\ 
Aol & 103 \\
Yandex & 76 \\ 
Ask & 14 \\ 
Search Encrypt & 0 \\
Baidu & 0 \\ 
PrivacyWall & 0 \\ 
\bottomrule
\end{tabular}

  \caption{Total number of searches participants made on twelve horizontal search engines during our observation window.}
  \label{tab:hor_search_count}
\end{table}

\subsubsection{Filtering Bing Activity}
\label{sec:filteringbing}

Microsoft has a rewards program that offers people monetary incentives to use services like Bing~\cite{microsoft_rewards}. One way that people can earn Microsoft rewards is to take quizzes on Bing. Answering a quiz question automatically submits a new query to Bing and appends a query parameter to the URL. This activity accounts for 28.9\% of Bing queries
and we assume that our sample over-represents people in the rewards program (43.8\% of participants who used Bing visited the Microsoft Rewards website a least once). Thus we filter out Bing queries that contain \texttt{Rewards}, \texttt{Quiz}, or \texttt{Gamification\_DailySet} in the query segment of the URL. 
This filter impacted 10.1\% of participants.

Additionally, we excluded one Bing user who made over 2000 Bing searches on a single day because we suspect that this participant was using automation to make searches. We did not exclude any Google users.

\subsection{Horizontal Search}
\label{sec:horizontal-search}


To study the market share of horizontal search engines, we developed a comprehensive list and identified queries to these engines in our datasets. We sourced horizontal search engines from regulatory reports~\cite{ftc2013,cma2020}, public usage statistics~\cite{statcounter-search-usa,statista-search-usa}, and ``search engine ballots'' that are mandatory in Europe due to prior regulatory actions against Google~\cite{warren-2019-ballot}. In total, our list contained 12 horizontal search engines.

To measure participants' usage of these horizontal search engines, we isolated URLs in participants' browsing history that matched the \textit{search schema} for these search engines (see \S\,\ref{sec:horizontal-schema} in the Appendix for details). We defined a search schema as the format used by a website to embed queries in its URLs, \eg \texttt{bing.com/search?q=QUERY} and \texttt{duckduckgo.com/?q=QUERY}. Of these 12 horizontal search engines, the top five by query volume are Google Search, Bing, DuckDuckGo, Yahoo Search, and Ecosia (see Table~\ref{tab:hor_search_count}). Subsequently, our analysis focused almost entirely on Google and Bing because they account for 97\% of our participants' searches on horizontal search engines.

Google and Microsoft each incorporate results from some of their vertical search engines into their horizontal search engine. To make our comparison between Google Search and Bing comprehensive and fair, we incorporate searches on a subset of each company's vertical search engines into the analysis of horizontal search behavior based on a list published by the UK CMA~\cite{cma2020_appendixC} (see \S\,\ref{sec:horizontal-schema} and Table~\ref{tab:hor_domains_and_url_parameter_key} in the Appendix for details). For example, we include searches on Google News and Bing News, but not searches on GMail, YouTube, Hotmail, or Microsoft Outlook when presenting results about Google Search and Bing's horizontal shares.

\subsection{Vertical Search}
\label{sec:vertical-search}


To analyze participants' search behavior on vertical search engines and compare it to their behavior on Google and Bing, we undertook the following steps:
\begin{enumerate}
  \item identify a mapping of websites to vertical segments (\eg Shopping, Travel, etc.),
  \item identify all searches that participants conducted on websites within the vertical segments, based on their browsing history, and
  \item divide participants' Google and Bing queries into the same vertical segments.
\end{enumerate}
This process enabled us to examine Google and Bing's market shares within each vertical segment relative to all other independent vertical search engines. We now describe each of these steps.

\subsubsection{Mapping Websites to Vertical Segments}
\label{sec:mapping-websites}

For this study we use the mapping of websites to vertical segments maintained by FortiGuard. FortiGuard is a vendor of cybersecurity software and their mapping is meant to help companies filter Internet traffic (\eg to block social media). Vallina \etal found that FortiGuard's mapping had the highest coverage of websites and the most accurate vertical segment labels compared to other vendors' mappings~\cite{vallina2020mis}. The FortiGuard mapping contains 90 vertical segments, which covered 157,792 (99.7\%) of the unique domains loaded by our participants; we highlight the most popular segments based on our participants' browsing history data in Table~\ref{tab:top_verticals_top_domains} in the Appendix.


\subsubsection{Identifying Searches on Websites}

To identify vertical search engines and participants' queries (if any) on these websites, we used a combination of manual and automated methods. First, we manually examined over 400 websites---a mix of the most popular websites overall and in specific vertical segments, sorted by participants' browsing history---to identify vertical search engines and their respective search schemas, \eg \texttt{amazon.com/s?k=QUERY}. This included examining 60 websites that we suspected might have non-keyword-based search functionality (\eg travel and restaurant reservation products) and 52 websites that required account registration (\eg social media). In total, these manually checked websites cover 71.1\% of all page loads in participants' browsing history.

Second, we built a web crawler that attempted to identify websites that supported search and their associated search schema. We instrumented the Chrome web browser to visit each website in our participants' browsing history and then applied the following two heuristics:
\begin{enumerate}
  \item The crawler tried to detect support for OpenSearch,\footnote{\url{https://developer.mozilla.org/en-US/docs/Web/OpenSearch}} which is a web standard that allows websites to programmatically expose their search functionality to web browsers. We used the search URL schema specified in the OpenSearch XML description to validate the effectiveness of our crawler.
  \item The crawler tried to locate an HTML \texttt{<input>} element where the keyword ``search'' appeared in (1) the \texttt{role} attribute of the \texttt{<form>} or (2) the \texttt{id}, \texttt{name}, \texttt{title}, \texttt{type}, or \texttt{class} properties of the \texttt{<input>} tag. If the crawler identified an input element matching these criteria, then it injected a unique query into the detected form, submitted the form, and then attempted to identify the query in the resulting URL.
  If the crawler found the query in the URL path (\eg \texttt{amazon.com/search/QUERY}) or in the URL parameters (\eg \texttt{amazon.com/s?k=QUERY}) we used this as the search schema for the website.
\end{enumerate}
These are the same heuristics used by prior work to investigate
search queries~\cite{kats-2022-pets}.
Overall, our crawler successfully visited 89.5\% of websites that appeared in participants' browsing history and parsed 96.7\% of the crawled websites' HTML. The crawler detected that 39.5\% of the parsed websites supported search functionality.

We validated the effectiveness of our crawler with both manual and automated checks. First, three authors manually reviewed the top 200 websites that the crawler identified as having search functionality. The authors identified the same search URL parameter as the crawler on 94\% of these sites. Second, on websites that had a syntactically-valid OpenSearch XML description, we compared the search URL parameter defined in the XML description file to that identified by our crawler. Our crawler agreed with the XML description file 95\% of the time.

Using the search schemas that we isolated for each of these websites, we separated the search and non-search URLs in participants' browsing history. From the 158,272 unique websites that appeared in participants' browsing history, we identified 48,978 (31.0\%) vertical search engines. Of the 7,848,032 page loads to these vertical search engines in our dataset, 293,401 (3.7\%) corresponded to searches. 


When we analyze participants' usage of vertical search engines, we include searches they performed on most Google and Microsoft products. For example, we include searches on GMail in the Web-based Email vertical segment and searches on YouTube in the Streaming Media segment (see \S\,\ref{sec:products} in the Appendix for further details).


\subsubsection{Assigning Google Queries to Vertical Segments}
\label{sec:assign-google}

To divide participants' Google queries into vertical segments---\eg Shopping queries, Travel queries, etc.---we performed a two stage classification process. First, we identified navigational queries and placed them in their own isolated vertical. Researchers have recognized that navigational queries are a distinct use case from \textit{informational queries}~\cite{broder2002taxonomy,jansen2008determining}, and regulators have noted that vertical search engines cannot answer navigational queries~\cite{cma2020_appendixP}.

We use Jansen \etal's rules-based approach to identify navigational queries~\cite{jansen2008determining} because it focuses on the content of the query, which allows us to apply it to both Google and Bing searches. We classify a search as navigational if the Jaro-Winkler similarity between a participant's query and the top-level domain of the next URL in their browsing history is $\geq 0.95$ \cite{cohen2003comparison}.
Overall, we identified 19,231 (7.1\%) navigational queries, which is similar to the 10--21\% navigational query rate identified by prior studies~\cite{jansen2008determining,teevan2011understanding}.

\begin{figure}[t]
  \centering
  \includegraphics[width = 1\linewidth]{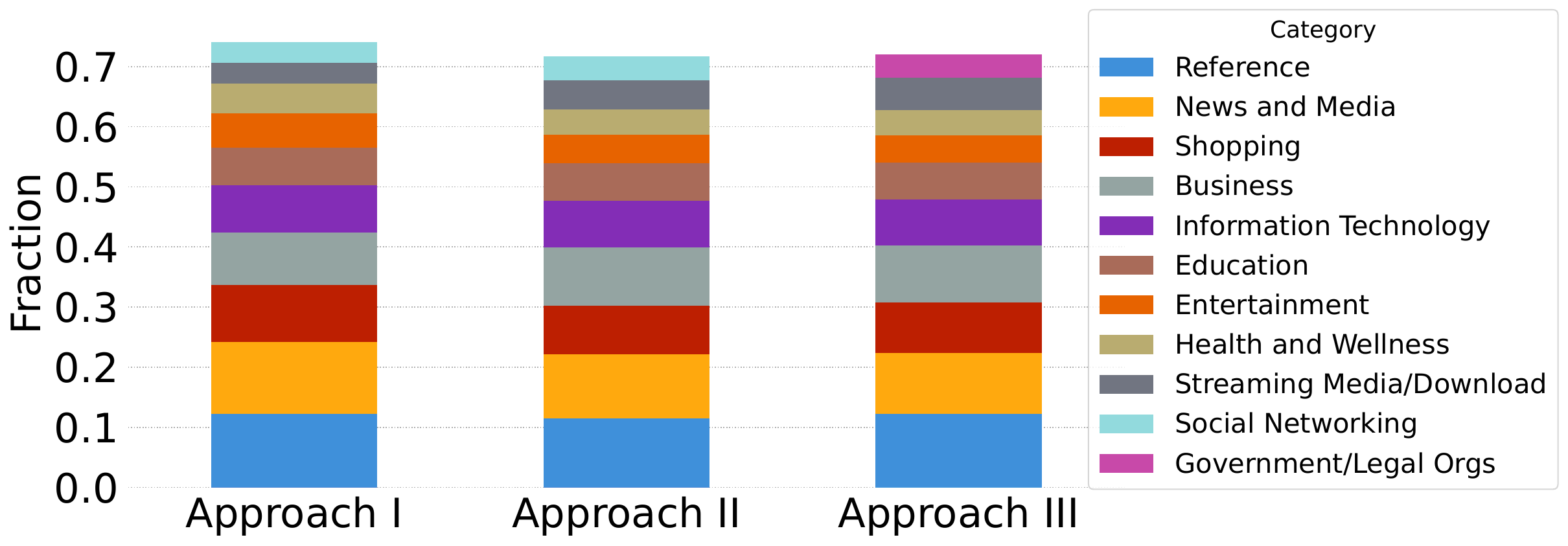}
  \caption{Shares of top ten vertical segments according to three different classification approaches. In each approach, SERPs with clicks are assigned the vertical segment of the clicked URL. SERPs without clicks are assigned a vertical segment based on the most frequently appearing segment (Approach I), segment distribution (II), and weighted segment distribution (III), respectively.}
  \label{fig:serp_methods}
\end{figure}

\begin{figure}[t]
\centering
\includegraphics[width = 1\linewidth]{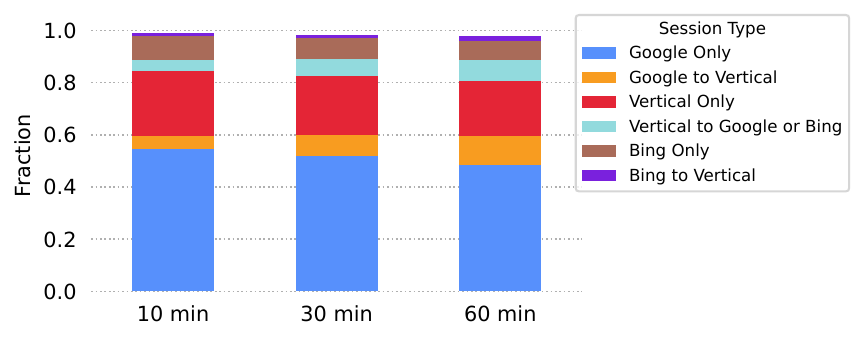}
\caption{Share of vertical sessions types under three rolling window sizes.}
\label{fig:plot_sessions_length_stack_bar}
\end{figure}

Second, we classified the remaining 251,831 (92.9\%) Google queries into the vertical segments from FortiGuard. If a participant clicked a link on a SERP, then we classified the query into the same vertical segment as the website in the clicked link. The intuition behind this strategy is that a person's intent when searching is revealed through their choice of result, as exemplified by their click. If a participant did not click any links on a SERP, then we treat the query as a weighted distribution over vertical segments. The distribution for a given query corresponds to the FortiGuard segments of the websites that are linked in the SERP, adjusted by weights that decrease geometrically with rank to account for attenuation in attention~\cite{papoutsaki2017gazer}. 

To assess the sensitivity of our SERP classification method, we evaluated two other SERP to vertical segment classification approaches: most frequent segment on the SERP and unweighted distribution over segments. Figure~\ref{fig:serp_methods} shows that the top ten vertical segments are extremely similar no matter what classification method is used. Furthermore, the three methods produce overall segment distributions (shown in Figure~\ref{fig:vertical_marketshare}) that are strongly correlated (Pearson $r \ge 0.914$, $p<0.001$ in all cases), which strongly suggests that our findings are robust to the choice of SERP classification method.

Note that we could not map 24.1\% of the Google SERPs in our corpus to vertical segments and we exclude them from all vertical analysis. This issue occurs because some Google SERPs contain results from Google's vertical search engines (\eg Google Images and Google Videos). It is unclear how to determine the appropriate vertical segment assignment for these searches, as the destinations of the links are not strong indicators of search intent. 
Given this constraint, our analysis of Google's share of vertical segments should be interpreted as a lower bound.

\subsubsection{Assigning Bing Queries to Vertical Segments}
\label{sec:assign-bing}

We perform the same two-stage classification process to assign Bing queries to vertical segments, with one important caveat: our browser extension did not collect snapshots of Bing SERPs. Therefore, we re-crawled participants' Bing queries on January 9--11, 2023 from a Boston IP address and extracted links from the SERPs using the open source \texttt{SearchParser} package.\footnote{\url{https://github.com/jlgleason/SearchParser}} Although the links on individual Bing SERPs likely differ from the ones participants viewed in 2020 (which prevents us from measuring clicks on Bing SERPs), we verified that the aggregate distribution over vertical segments for a fixed sample of 1,000 Google queries was similar between late 2020 and early 2023.\footnote{Specifically, the Jensen-Shannon distance ($JS$) between the aggregate 2020 and 2023 distributions was 0.1. As a reference, $JS([0.5, 0.3, 0.2], [0.55, 0.35, 0.1]) = 0.1$.}
Thus, we treat Bing SERPs as weighted distributions over vertical segments and rely on the assumption that this represents aggregate Bing search behavior from late 2020 with high fidelity. 

As with Google, we exclude 15.6\% of Bing SERPs in our corpus from our analysis of vertical segments because they contain results from Microsoft's vertical search engines that do not clearly map to a vertical segment (primarily Bing Images and Bing Videos).

\begin{figure*}[t]
  \centering
  \begin{subfigure}[t]{0.33\linewidth}
    \includegraphics[width=\linewidth]{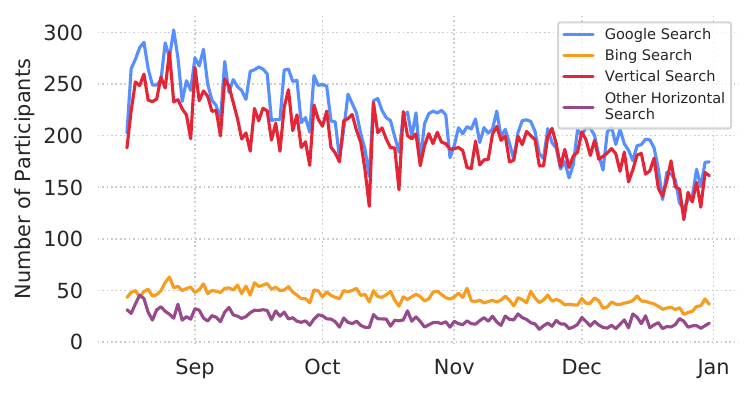}
    \caption{Participants per Day}
    \label{fig:users_vs_time}      
  \end{subfigure}
  \hfill
  \begin{subfigure}[t]{0.33\linewidth}
    \includegraphics[width=\linewidth]{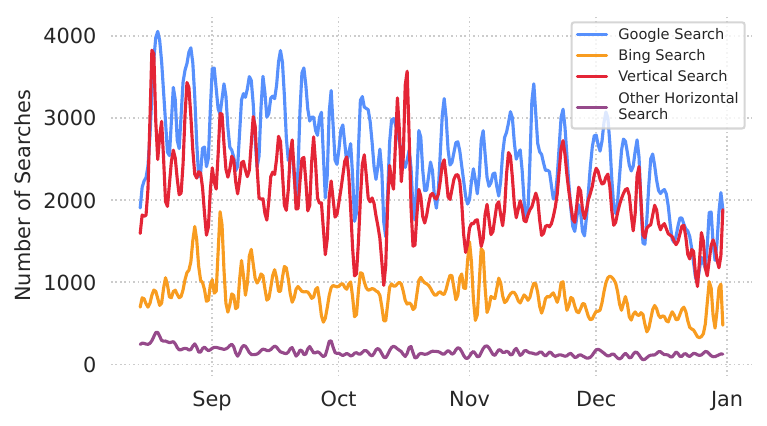}
    \caption{Searches per Day}
    \label{fig:search_volume_vs_time}
  \end{subfigure}
  \begin{subfigure}[t]{0.33\linewidth}
    \includegraphics[width=\linewidth]{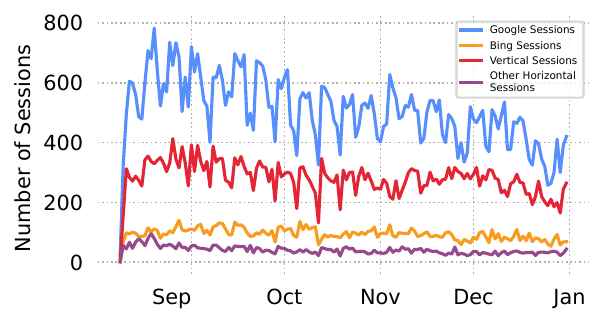}
    \caption{Sessions per Day}
    \label{fig:sessions_per_day_vs_time}
  \end{subfigure}
  \caption{\textbf{On 86.3\% of the days in our five month observation window, Google Search was used more than all vertical search engines combined.} Figure~\ref{fig:users_vs_time} presents the number of participants that conducted at least one search in each category (\eg Google Search, Bing, another horizontal search engine, or any vertical search engine) per day, Figure~\ref{fig:search_volume_vs_time} presents the total number searches per day per category, and Figure~\ref{fig:sessions_per_day_vs_time} presents the total number of sessions per day per category. Google's vertical search engines (\eg GMail, YouTube, etc.) are counted among all other vertical search engines in these figures. Participants' usage trends were steady over time after accounting for natural attrition among participants. Google Search was consistently used more than competing horizontal search engines, although Bing users conducted more searches per person per day than Google users.}
  \label{fig:time}
\end{figure*}

\subsection{Search Sessions}
\label{sec:search-sessions}

One goal of our study is to assess Google Search and Bing's gatekeeping power by examining where our participants begin and end information seeking tasks. If participants predominantly begin seeking information via Google Search, for example, this grants Google the power to steer participants to subsequent vertical search engines (owned by third-parties or Google itself) where they may refine their queries.

To investigate gatekeeping power we examine participants' propensity to switch between search engines during a single \textit{search session}. Like prior work, we define a search session as searches that occur within a rolling 30 minute window of each other~\cite{jansen2007defining,white2009characterizing,hassan2014struggling}. 77\% of Google searches and 88\% of Bing searches in our dataset have an inter-arrival time under 30 minutes, which further motivates this threshold.

In this study, we examine \textit{vertical search sessions}, which include searches made on Google Search, other Google products (\eg YouTube, GMail, and Drive), Bing, other Microsoft products (\eg Bing News, Shopping, and Travel), and/or independent vertical search engines.
We represent vertical search sessions as a distribution over verticals in which each search receives equal weight. Specifically, the vertical distribution for a session is $\mathbf{s} = \frac{1}{n}*\sum_{j=1}^{n}\mathbf{c_{j}}$, where $n$ is the number of searches in a session and $\mathbf{c_{j}}$ is the vertical distribution for search $j$ in the session. Using this approach we constructed 131,802 vertical search sessions, 82.2\% of which include only a single search engine.

To assess the sensitivity of our assignment of search sessions to vertical segments, we repeated the assignment procedure as we varied the rolling window size from 10 to 60 minutes in increments of 10 minutes. Figure~\ref{fig:plot_sessions_length_stack_bar} shows that the distributions of session types are very similar regardless of the window size. Specifically, we found that the vertical segment distributions were strongly correlated (Pearson $r \ge 0.966$, $p<0.001$ in all cases), which validated that our assignment approach is robust.

\section{Analysis}
\label{sec:analysis}

In this section we present the results of our analysis, first by focusing on horizontal search engines, then by examining all search engines used by our participants.

\begin{table}[t]
  \centering
  \footnotesize
  \begin{tabular}{lrrr}
    \toprule
    \textbf{Search Engine} & \textbf{\% Participants} & \textbf{\% Searches}  & \textbf{\% Sessions}   \\
    \midrule
    Google Search & 79.3 & 71.8 & 78.9 \\ 
    Bing          & 11.2  & 23.8 & 14.8 \\
    DuckDuckGo    &  3.8 &  2.0  & 2.9 \\
    Yahoo Search  &  4.8 &  2.0  & 2.9\\
    Ecosia        &  0.8 &  0.4  & 0.5\\
    \bottomrule
  \end{tabular}
  \caption{\textbf{Google Search has the greatest market share among our participants when compared to other horizontal search engines.} This is true whether we consider participants' preferred search engines (\ie where they did the majority of their searches), overall search volume, or search sessions.}
  \label{tab:hor_share}
\end{table}

\begin{figure*}[t]
  \centering
  \includegraphics[width =1.0 \textwidth] {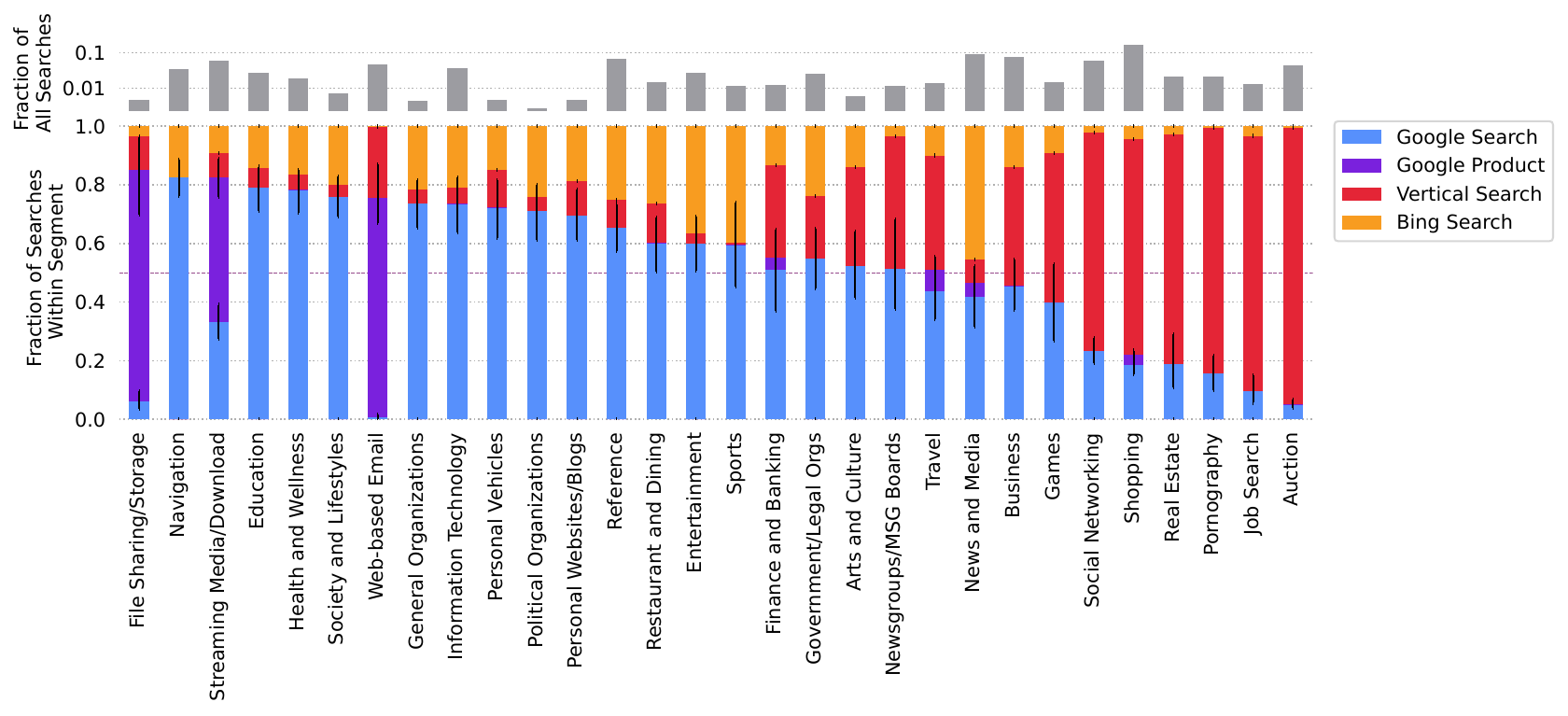}
  \caption{\textbf{Google receives $>$ 50\% of searches in \verticalsvolume out of 30 vertical segments.} We examine the fraction of searches in each of the top 30 vertical segments that occurred on Google Search, another Google product, Bing, or an independent vertical search engine, and perform 1000 bootstrap replications to compute 95\% confidence intervals. Five of the verticals where Google Search receives $>$ 50\% of searches are within the confidence interval. The top inset shows the overall distribution of searches.}
  \label{fig:vertical_marketshare}
\end{figure*}

\subsection{Horizontal Search}

Table~\ref{tab:hor_share} presents the market share of horizontal search engines, measured in terms of participants and overall search volume. Some of our participants use multiple horizontal search engines---we assume that the horizontal search engine a participant used greater than 50\% of the time is their preferred search engine.\footnote{The percentages in the `\% Participants' column of Table~\ref{tab:hor_share} only sum to 99.9\% because a small number of participants did not use any horizontal search engine greater than 50\% of the time.}
Given this threshold, 79.3\% of our participants primarily used Google Search. In terms of overall search volume, Google Search received 71.8\% of participants' searches on horizontal search engines, which is within the 61--80\% range of publicly available estimates for Google Search's share of desktop searches in the US~\cite{statcounter-search-usa,statista-search-usa}. Our estimate of Bing's share based on search volume (after removing the quiz reward related queries) is also within the publicly available estimates (12--25\%) and our estimate of Yahoo Search's share is slightly below (2.0 \% versus 4--11\%). In terms of horizontal search sessions, 78.9\% of our participants sessions began at Google Search and 14.8\% began at Bing. Given the low usage rates of DuckDuckGo, Yahoo Search, and Ecosia among our participants, we focus our attention on Google Search and Bing in the remainder of our analysis.

Figure~\ref{fig:time} presents our participants' search volume over time---in terms of participants per day, total searches per day, and sessions per day---divided between Google Search, Bing, other horizontal search engines, and vertical search engines (which includes vertical search engines like GMail, Outlook, and YouTube operated by Google and Microsoft). We draw three observations from Figure~\ref{fig:time}.

First, the relative usage of the four categories is consistent throughout our observation window. Google Search is most used, regardless of how we quantify search behavior per day, followed by vertical search engines, Bing, and other horizontal search engines (DuckDuckGo, etc.). 

Second, the consistency we observe over time suggests that our sample was not impacted by any exogenous events. In particular, browser and operating system vendors sometimes attempt to unilaterally change users' default search engine~\cite{warren-2020-bing,warren-2021-edge}, but we observe no significant changes in overall or individual-level horizontal search engine usage during our observation window. We do observe a gradual decline in all metrics over the time that is due to natural attrition of participants over the study period.

\begin{figure*}[t]
  \centering
  \includegraphics[width =1.0 \textwidth] {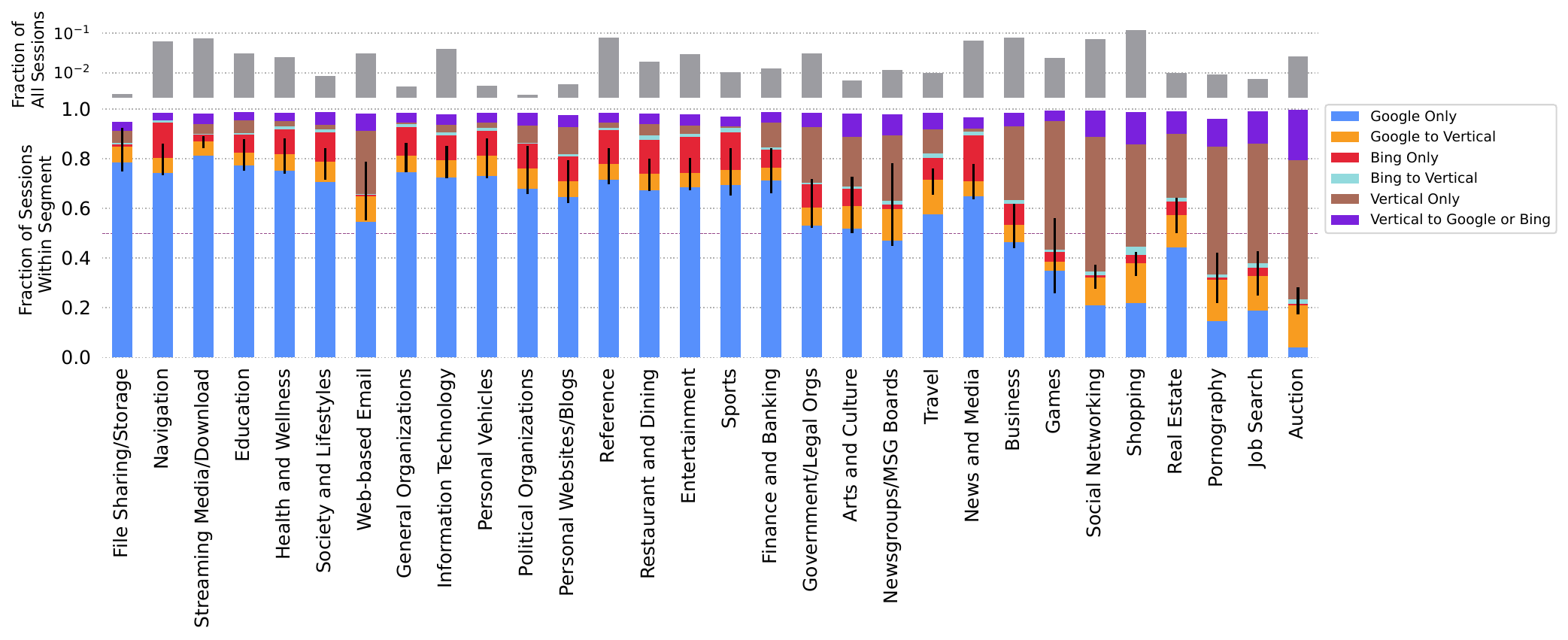}
  \caption{\textbf{$>$ 50\% of participants' search sessions begin on a Google product (or solely involve a Google product) in \verticalssession out of 30 vertical segments.} We examine the fraction of search sessions in the top 30 vertical segments including: only searches on Google products, only on Bing, only on vertical search engines, or searches on two of the three. In the latter case, we divide the search sessions based on where the initial search in each session occurred. We compute 95\% confidence intervals for the sum of `Google Only' and `Google to Vertical' categories using 1000 bootstrap replications. The top inset shows participants' session distribution.}
  \label{fig:google_gatekeeper_sessions}
\end{figure*}

Third, there are differences in participants' usage of Google Search and Bing. Comparing Figure~\ref{fig:users_vs_time} and Figure~\ref{fig:search_volume_vs_time}, we see that Bing received more searches per participant than Google Search. Considering only Google Search and Bing, 73.4\% of participants in our sample only used Google Search and 3.2\% only used Bing. Among these participants, Google users averaged 7.5 searches per day while Bing users averaged 9.9 searches per day.\footnote{Participants conducted an average of 12.5 Bing searches per day before we filtered quiz searches out, see \S\,\ref{sec:filteringbing}.} We hypothesize that Bing users may be more active due to Microsoft's Bing Rewards program, which pays people in the form of gift cards for using Bing. That said, despite Bing Rewards existing since 2010 and being popular among a subset of our participants,
Google Search still receives more than two times Bing's share.

We find that participants' rarely switch between Google Search and Bing at short timescales. Of the 23.3\% of our participants that used each of Google Search and Bing at least once during our observation window, 14\% used Google for the majority of their searches versus 9.3\% for Bing.  Out of 99,248 horizontal search sessions in our dataset, only 1.8\% contain at least one Google Search and at least one Bing search. This suggests that participants who use Google Search and Bing rarely abandon queries on one search engine and retry them on the other---instead, participants may use the two engines for different types of queries or in different contexts.


\subsection{Vertical Search}

\subsubsection{Market Share}

Figure~\ref{fig:vertical_marketshare} presents the fraction of participants' searches within thirty vertical segments stratified by where they occurred: on Google Search, on a non-Search Google product, on Bing, or on an independent vertical search engine. We compute 95\% confidence intervals using the percentile bootstrap over participants with 1,000 replications. Figure~\ref{fig:vertical_marketshare} focuses on the thirty most popular verticals in our dataset, which collectively account for 94.1\% of participants' vertical searches. The volume of searches in each vertical segment is shown in the upper portion of Figure~\ref{fig:vertical_marketshare}. We sort the vertical segments along the x-axis based on Google's share of search volume, computed as the sum of Google Search and Google product search volume within each segment.

Figure~\ref{fig:vertical_marketshare} shows that Google products receive greater than 50\% of search volume in \verticalsvolume vertical segments, although six are within the confidence interval. Google products receive greater than 50\% of search volume in many informational segments, such as Health and Wellness, Entertainment, and Reference, despite major websites in these segments having their own search functionality. Searches on YouTube, Google Drive, and GMail account for Google's share in the Streaming Media, File Sharing/Storage, and Web-based Email segments, respectively.\footnote{Our results in the Web-based Email and File Sharing/Storage verticals should be interpreted with caution: because Hotmail, Outlook, and OneDrive do not encode queries in their URLs, we are unable to measure searches on these services.} Google receives over 80\% of participants' navigational queries and no vertical search engines appear in this segment.

In segments where Google products receive less than 50\% of search traffic, eBay receives the most queries in the Auction segment, Zillow in Real Estate, Indeed in Job Search, Twitter and Facebook in Social Networking, and Amazon in Shopping.
Out of the top 30 segments, Bing matches Google's share in only one segment: News and Media. We hypothesize that this may be driven by links to breaking news that Microsoft includes on MSN, the Bing homepage, and in Windows~\cite{win10-bing-alerts}.


\subsubsection{Gatekeeping Power}

We investigated Google and Bing's role as gatekeeper by constructing and analyzing vertical search sessions. Our goal is to capture participants' proclivity for switching between Google and Microsoft-owned search engines and vertical search engines, as well as understand whether participants start their information seeking tasks on Google and Microsoft products or on vertical search engines.

Figure~\ref{fig:google_gatekeeper_sessions} presents the fraction of vertical search sessions in each vertical segment that included only searches on Google products, only searches on Bing, only searches on vertical search engines, or sessions that include searches on two of the three, broken down based on where the first search in each session was initiated. To make the figure legible, we omit sessions that included searches from all three (0.7\% of total sessions) and sessions that only include Google products and Bing (1.0\% of total sessions). We compute 95\% confidence intervals using the percentile bootstrap over participants with 1,000 replications. Although Figure~\ref{fig:google_gatekeeper_sessions} contains six session types, we only computed a confidence interval for the most consequential: the sum of `Google Only' and `Google to Vertical' sessions. Figure~\ref{fig:google_gatekeeper_sessions} retains the same sort order of segments along the x-axis as Figure~\ref{fig:vertical_marketshare}. Moreover, the thirty categories in Figure~\ref{fig:google_gatekeeper_sessions} collectively account for 92.4\% of participants' vertical sessions.

Overall, 16.2\% of vertical sessions in Figure~\ref{fig:google_gatekeeper_sessions} include switching, versus 1.0\% of vertical sessions when we only consider vertical sessions with Google products and Bing. These observations make intuitive sense: as horizontal search engines, Google Search and Bing are obviously substitutable, so there is relatively little incentive for people to switch between them at short time scales. In contrast, vertical search engines are much less substitutable. For example, Google Search can help you find a specific item for sale or a particular retailer, but the best way to see a retailers' full inventory in a legible format is to search on their own website. Furthermore, some information is not public on the web and is thus inaccessible from Google Search, such as private posts on social media websites that can only be surfaced by using their native search interfaces.

The results in Figure~\ref{fig:google_gatekeeper_sessions} demonstrate that Google has significant gatekeeping power. In \verticalssession of the top 30 segments, sessions begin on Google products (or only include searches on Google products) at least 50\% of the time.

We observe that in many verticals, Google's proportion of the market grows substantially when we shift from the granularity of individual searches (Figure~\ref{fig:vertical_marketshare}) to search sessions (Figure~\ref{fig:google_gatekeeper_sessions}). For example, in the Real Estate vertical, Google Search receives roughly 20\% of individual searches, but 57.3\% of sessions begin on a Google product. We make similar observations for the Finance and Banking and Travel verticals. Bing's share of the News and Media vertical shrinks when we change the unit of analysis from individual searches to search sessions because long sequences of news queries get collapsed into single sessions.

These results highlight how our conception of market share, and thus market power, in online search may shift depending on how we account for user behavior---in this case, the tendency for people to conduct multiple searches in rapid succession within task-oriented sessions. Further, these findings demonstrate the extent to which vertical search engines are reliant on Google products for traffic.


\section{Discussion}
\label{sec:discussion}

We conclude by discussing the implications of our findings and the limitations of our study.

\subsection{Search Market Dynamics}
\label{sec:search_market_dynamics}

Defining the boundaries of relevant search markets lies at the core of some of the most consequential antitrust litigation against Google, and has the potential to shape the future of the digital economy. Our study offers new methods and empirically-derived insights in the market(s) for online search, beyond the existing focus on aggregate usage measures and high-level arguments about corporate revenue~\cite{harkrider-2020-dennys}. We provide a potential basis for more granular, vertical segmentation of markets for online search and a better understanding of Google's gatekeeping position.

We observe that our participants exhibit entrenched behavioral patterns with respect to online search. 79.3\% of our participants preferred Google Search over other horizontal search engines, and this preference was stable over time. On most days, Google Search received more searches than all other vertical search engines combined---a tally that even includes searches on Google products like GMail and YouTube.

Our analysis supports the concerns of market participants, regulators, legal scholars, and journalists who posit that Google may have the power to leverage its dominance in the horizontal search market to also dominate specific vertical segments by, for example, affording preferential placement to Google's vertical search products in SERPs~\cite{house2020,ec_shopping_full,cma2020,khan-2019-clr,jeffries-2020-google,gleason2023google}. Over 80\% of our participants' navigational queries were to Google Search, which gives Google an immediate advantage over all vertical competitors since they cannot answer these queries. In \verticalssession out of 30 vertical segments, participants' search sessions began with a Google product greater than 50\% of the time (and only 13.9\% of search sessions included transitions between a Google product and a non-Google vertical search engine across all verticals). This gatekeeper position grants Google enormous power to collect data about the preferences of Internet users, as well as steer participants towards vertical search engines of Google's choosing.

Among the nine vertical segments where Google commanded less than 50\% of search volume, many are dominated by other major tech platforms (\eg Amazon, ebay, Indeed, Facebook, Twitter, and Zillow). These are segments that Google has entered or attempted to enter in the past (\eg through the introduction of Google+, Google Shopping, and Google for Jobs). The Shopping vertical has been the focus of previous regulatory actions against Google~\cite{ec_shopping_full} and travel companies are advocating for similar regulatory intervention in Europe~\cite{lomas-2020-travel}. Our analysis supports regulatory scrutiny of Google's actions in these verticals. For example, we observe that Google receives 51\% ($+$15\%, $-$13\%) of searches in the Travel vertical among our participants, but Google could potentially tip the scales unambiguously in their own favor by prioritizing Google Flights and Google Hotels in SERPs while demoting other travel companies.

One of the vertical segments that Google does not dominate---Pornography---may have been intentionally ceded by the company. Google Search filters pornography from search results by default unless users disable the Safe Search feature, and Google prohibits pornography on other platforms they own (\eg YouTube and the Play Store).

Understanding why people exhibit entrenched online search behaviors requires further study. Google claims that people prefer their search engine because it offers the highest-quality results~\cite{walker-2020-doj}. It is also plausible, however, that bundling of search engines with hardware and software~\cite{ec_android}, as well as default effects~\cite{lyons-2020-mozilla}, may ossify peoples' search behavior.

Although generative artificial intelligence (AI) models are rapidly altering the affordances of search engines, it is unclear whether they will have lasting impacts on search market shares. As of May 2023, Google Search and Bing both integrate chat-style AI based on large language models~\cite{peters-2023-googleio}. While Bing was first to adopt this technology, initial speculation that this would cause Bing to take market share from Google Search does not appear to be coming true~\cite{dotan-2023-wsj}. This could be because defaults (which favor Google Search) are sticky, users' initial excitement for these technologies has waned, or simply that overcoming entrenched human behaviors is hard for upstarts. More broadly, online search startups are failing~\cite{pierce-2023-neeva} and business relationships between Google Search and dominant firms are not changing~\cite{roth-2023-samsung}.

\subsection{Assessing Monopoly Power}
\label{sec:monopoly}

In this study we focus on market definitions and shares, which are preliminary questions in establishing whether a company holds monopoly power. When assessing whether a company's market power amounts to monopoly power, courts conduct holistic assessments that may, for example, include considerations of market entry barriers. This holistic assessment remains beyond the scope of this study. Moreover, a finding of monopoly power only opens the door to antitrust enforcement; it is necessary, but not sufficient to support an antitrust claim. Standing antitrust doctrine requires specific anticompetitive behavior and harm in the form of consumer welfare losses to support a claim~\cite{1979reiter,2018ohio,areeda-antitrust}.

A more granular, vertical segmentation of markets for online search does not suggest that the effects of Google's monopoly power are restricted to the particular vertical in which it holds a significant share. Google may leverage its monopoly power in one vertical or its gatekeeping power over navigational queries to exert power in vertical segments in which it holds a comparatively small share. Additionally, our study only examines one side of a multi-sided platform, and does not address Google's potential to leverage its position in the online advertising market~\cite{us_texas}.

Scholars have criticized the reliance on market shares as an indicator of market power and, more generally, the definition of relevant markets, due to the inherent challenges and uncertainties associated with that practice~\cite{hovencamp-2022-cblr}. Instead, they have suggested to rely on direct evidence of harm. Our study takes no position on this issue, but acknowledges that regulators, enforcers, and courts continue to consider market shares and require market definitions~\cite{2018ohio}. 

\subsection{Limitations}
\label{sec:limitations}

Our study has several limitations. Our data is constrained to a sample of online activity from US-based individuals on the desktop platform and our participants may not be perfectly representative. Participants who agreed to install our browser extension expressed slightly higher usage and trust in Google Search than participants who did not agree to install our extension.
That said, many of our participants engaged with the Microsoft Rewards Program, which incentivizes users to search on Bing.
Publicly available estimates indicate that Google Search has greater than 90\% horizontal market share in mobile and non-US markets (\eg the UK and Europe)~\cite{statcounter-search-global}. This suggests that our results should be treated as a lower-bound on Google's dominance across horizontal and vertical market segments.

To respect participants' privacy, our browser extension did not collect data from incognito browsing windows. It is unclear how often people used incognito mode during our data collection period, although we note that we did capture a significant amount of potentially sensitive browsing activity (\eg searching for and viewing pornography). This suggests that at least some participants did not use---or did not consistently use---this functionality.

Our analysis depends on specific data operationalization choices that could impact our conclusions. We only consider the first click on a SERP when assigning it a category; taking additional clicks into account could alter the distribution of query volume across segments.

Finally, there may be false negatives in our detection of participants' searches on independent vertical search engines. 
However, we manually validated that we correctly identified the search schemas on websites that account for 71.1\% of all page loads in participants' browsing history, so the impact of potential false negatives is constrained. A related issue is that Hotmail, Outlook, and OneDrive do not include queries in their URLs, which prevents us from tabulating searches on these services in the Web-based Email and File Sharing segments, respectively. 

\section{Broader Perspective}
\label{sec:broader}

This study was approved by the Northeastern IRB under protocol \#20-03-4. All participants consented to data collection (see \S~\ref{sec:informed-consent}) and were compensated. The total amount we paid to YouGov to administer our survey and compensate participants was \$78,000. Participants were free to leave our study at any time. Our browser extension used TLS to protect data in transit and uninstalled itself at the end of the study period. Participant data was stored on a siloed server that was only accessible to personnel approved by the IRB. 

We do not foresee any negative societal impacts of this study or risks to study participants. The nature of the data we collected from participants precludes deidentification. Thus, in accordance with our protocol, we only present aggregated results in this manuscript and we will not be making identifiable data from this study publicly available.

\section*{Acknowledgements}    
The collection of data used in this study was funded in part by the Anti-Defamation League, the Russell Sage Foundation, and the Democracy Fund. This research was supported in part by NSF grant IIS-1910064. Any opinions, findings, and conclusions or recommendations expressed in this material are those of the authors and do not necessarily reflect the views of the funders.

\bibliography{bibtex-all}
\newcommand{\answerYes}[1]{\textcolor{blue}{#1}} 
\newcommand{\answerNo}[1]{\textcolor{teal}{#1}} 
\newcommand{\answerNA}[1]{\textcolor{gray}{#1}} 
\newcommand{\answerTODO}[1]{\textcolor{red}{#1}} 

\section{Appendix}

\subsection{Ethics Checklist}

\begin{enumerate}

  \item For most authors...
  \begin{enumerate}
      \item  Would answering this research question advance science without violating social contracts, such as violating privacy norms, perpetuating unfair profiling, exacerbating the socio-economic divide, or implying disrespect to societies or cultures?
      \answerYes{Yes}
    \item Do your main claims in the abstract and introduction accurately reflect the paper's contributions and scope?
    \answerYes{Yes}
     \item Do you clarify how the proposed methodological approach is appropriate for the claims made? 
     \answerYes{Yes, see \S~\ref{sec:methods}.}
     \item Do you clarify what are possible artifacts in the data used, given population-specific distributions?
      \answerYes{Yes, see \S~\ref{sec:methods:participants}.}
    \item Did you describe the limitations of your work?
      \answerYes{Yes, see \S~\ref{sec:limitations}.}
    \item Did you discuss any potential negative societal impacts of your work?
      \answerYes{Yes, see \S~\ref{sec:broader}.}
      \item Did you discuss any potential misuse of your work?
      \answerNA{NA}
      \item Did you describe steps taken to prevent or mitigate potential negative outcomes of the research, such as data and model documentation, data anonymization, responsible release, access control, and the reproducibility of findings?
      \answerYes{Yes, see \S~\ref{sec:broader}.}
    \item Have you read the ethics review guidelines and ensured that your paper conforms to them?
      \answerYes{Yes}
  \end{enumerate}
  
  \item Additionally, if your study involves hypotheses testing...
  \begin{enumerate}
    \item Did you clearly state the assumptions underlying all theoretical results?
    \answerNA{NA}
    \item Have you provided justifications for all theoretical results?
    \answerNA{NA}
    \item Did you discuss competing hypotheses or theories that might challenge or complement your theoretical results?
    \answerNA{NA}
    \item Have you considered alternative mechanisms or explanations that might account for the same outcomes observed in your study?
    \answerNA{NA}
    \item Did you address potential biases or limitations in your theoretical framework?
    \answerNA{NA}
    \item Have you related your theoretical results to the existing literature in social science?
    \answerNA{NA}
    \item Did you discuss the implications of your theoretical results for policy, practice, or further research in the social science domain?
    \answerNA{NA}
  \end{enumerate}
  
  \item Additionally, if you are including theoretical proofs...
  \begin{enumerate}
    \item Did you state the full set of assumptions of all theoretical results?
      \answerNA{NA}
    \item Did you include complete proofs of all theoretical results?
      \answerNA{NA}
  \end{enumerate}
  
  \item Additionally, if you ran machine learning experiments...
  \begin{enumerate}
    \item Did you include the code, data, and instructions needed to reproduce the main experimental results (either in the supplemental material or as a URL)?
      \answerNA{NA}
    \item Did you specify all the training details (e.g., data splits, hyperparameters, how they were chosen)?
      \answerNA{NA}
    \item Did you report error bars (e.g., with respect to the random seed after running experiments multiple times)?
      \answerNA{NA}
    \item Did you include the total amount of compute and the type of resources used (e.g., type of GPUs, internal cluster, or cloud provider)?
      \answerNA{NA}
    \item Do you justify how the proposed evaluation is sufficient and appropriate to the claims made? 
      \answerNA{NA}
    \item Do you discuss what is ``the cost'' of misclassification and fault (in)tolerance?
      \answerNA{NA}
    
  \end{enumerate}
  
  \item Additionally, if you are using existing assets (e.g., code, data, models) or curating/releasing new assets, \textbf{without compromising anonymity}...
  \begin{enumerate}
    \item If your work uses existing assets, did you cite the creators?
    \answerNA{NA}
    \item Did you mention the license of the assets?
    \answerNA{NA}
    \item Did you include any new assets in the supplemental material or as a URL?
    \answerNA{NA}
    \item Did you discuss whether and how consent was obtained from people whose data you're using/curating?
    \answerYes{Yes, see \S~\ref{sec:broader}.}
    \item Did you discuss whether the data you are using/curating contains personally identifiable information or offensive content?
    \answerYes{Yes, see \S~\ref{sec:broader}.}
    \item If you are curating or releasing new datasets, did you discuss how you intend to make your datasets FAIR?
    \answerNA{NA}
    \item If you are curating or releasing new datasets, did you create a Datasheet for the Dataset? 
    \answerNA{NA}
  \end{enumerate}
  
  \item Additionally, if you used crowdsourcing or conducted research with human subjects, \textbf{without compromising anonymity}...
  \begin{enumerate}
    \item Did you include the full text of instructions given to participants and screenshots?
    \answerYes{Yes, see \S~\ref{sec:broader} and \S~\ref{sec:informed-consent}.}
    \item Did you describe any potential participant risks, with mentions of Institutional Review Board (IRB) approvals?
      \answerYes{Yes, see \S~\ref{sec:broader}.}
    \item Did you include the estimated hourly wage paid to participants and the total amount spent on participant compensation?
      \answerYes{Yes, see \S~\ref{sec:broader}. Note that YouGov handled individual participant compensation and we are not privy to per participant wages.}
    \item Did you discuss how data is stored, shared, and deidentified?
      \answerYes{Yes, see \S~\ref{sec:broader}.}
  \end{enumerate}
  
  \end{enumerate}

\subsection{Browser Extension Informed Consent}
\label{sec:informed-consent}

\noindent Welcome to the study!\\

\noindent This extension implements a user study being conducted by researchers at Northeastern University, Dartmouth, Princeton, and University of Exeter. If you choose to participate, this browser extension will confidentially collect four types of data from your browser.\\

\noindent 1. Metadata for web browsing (\eg URL visited with time of visit), exposure to embedded URLs on websites (\eg YouTube videos), and interactions with websites (\eg clicks and video viewing time). This data is collected until the study is completed.\\

\noindent 2. Copies of the HTML seen on specific sites: Google Search, Google News, YouTube, Facebook Newsfeed, and Twitter Feed. We remove all identifying information before it leaves the browser. This confidential data is collected until the study is completed.\\

\noindent 3. Browsing history, Google and YouTube account histories (\eg searches, comments, clicks), and online advertising preferences (Google, Bluekai, Facebook). This data is initially collected for the year prior to the installation of our browser extension, and we then check these sources once every two weeks to collect updates until the study is completed.\\

\noindent 4. Snapshots of selected URLs from your browser. For each URL, the extension saves a copy of the HTML that renders, effectively capturing what you would have seen had you visited that website yourself. Once per week we conduct searches on Google Search, Google News, YouTube, and Twitter, and collect the current frontpage of Google News, YouTube, and Twitter. These web page visits will occur in the background and will not affect the normal functioning of your browser. There is a theoretical risk of ``profile pollution'' -- that this extension will impact your online profiles, i.e., ``pollute'' them with actions that you did not take. To mitigate this risk, the extension will only visit content that is benign and will only execute searches for general terms. Our previous work has found that historical information of this kind has minimal impact on online services.\\

\noindent Additionally, if you choose to participate, you will be asked to take a survey in which we ask you several questions about your demographics, web usage, and media preferences. These data, as well as those mentioned above, will be used to analyze the correlations between your online behavior and your interest profiles.\\

\noindent After the study is complete on December 31, 2020, the extension will uninstall itself. All data collected will be kept strictly confidential and used for research purposes only. We will not share your responses with anyone who is not involved in this research.\\

\noindent You must be at least 18 years old to take part in this study. The decision to participate in this research project is voluntary. You do not have to participant and you can refuse to participate. Even if you begin our experiment, you can stop at any time. You may request that we delete all data collected from your web browser at any time.\\

\noindent We have minimized the risks. We are collecting basic demographic information, information about your internet habits, and copies of web pages that you visit. To the greatest extent possible, information that identifies you will be removed from all collected web data.\\

\noindent Your role in this study is confidential. However, because of the nature of electronic systems, it is possible, though unlikely, that respondents could be identified by some electronic record associated with the response. Neither the researchers nor anyone involved with this study will be collecting those data. Any reports or publications based on this research will use only aggregate data and will not identify you or any individual as being affiliated with this project.\\
\begin{table}[t]
    \resizebox{\linewidth}{!}{%
    \begin{tabular}{lrrrr}
    \toprule
    {} & {} &  \textbf{Ext.} &  \textbf{No Ext.} &  \textbf{Diff.} \\
    \midrule
    \textbf{Gender} & Male                 &       0.47 &          0.47 &        0.01 \\
    & Female               &       0.53 &          0.53 &       -0.01 \\
    \midrule
    \textbf{Race} & White                &       0.74 &          0.68 &        0.06 \\
    & Black                &       0.12 &          0.13 &       -0.01 \\
    & Hispanic             &       0.08 &          0.09 &       -0.01 \\
    & Asian                &       0.02 &          0.06 &       -0.04 \\
    & Native American      &       0.01 &          0.01 &        0.00 \\
    \midrule
    \textbf{Employment} & Full-time            &       0.36 &          0.35 &        0.01 \\
    & Part-time            &       0.13 &          0.09 &        0.04 \\
    & Unemployed           &       0.07 &          0.09 &       -0.02 \\
    & Retired              &       0.20 &          0.22 &       -0.02 \\
    & Permanently disabled &       0.09 &          0.07 &        0.02 \\
    \midrule
    \textbf{Education} & High school graduate &       0.25 &          0.39 &       -0.14 \\
    & Some college         &       0.25 &          0.19 &        0.06 \\
    & 2-year               &       0.11 &          0.12 &       -0.01 \\
    & 4-year               &       0.25 &          0.17 &        0.08 \\
    & Post-grad            &       0.13 &          0.10 &        0.03 \\
    \midrule
    \textbf{Marital Status} & Married              &       0.43 &          0.46 &       -0.04 \\
    & Divorced             &       0.12 &          0.13 &       -0.01 \\
    & Never married        &       0.32 &          0.29 &        0.03 \\
    \midrule
    \textbf{Party Identification} & Democrat             &       0.42 &          0.31 &        0.11 \\
    & Republican           &       0.20 &          0.30 &       -0.10 \\
    & Independent          &       0.28 &          0.29 &       -0.01 \\
    \midrule
    \textbf{Age} & 18-34                &       0.26 &          0.24 &        0.02 \\
    & 35-54                &       0.29 &          0.32 &       -0.03 \\
    & 55-64                &       0.20 &          0.19 &        0.01 \\
    & 65+                  &       0.25 &          0.25 &        0.00 \\
    \midrule
    \textbf{Sample Size (weighted)} & {} & 790 & 1210 & {} \\
    \bottomrule
\end{tabular}
    
    }
    \caption{Comparison of weighted demographics in the extension and no extension samples.}
    \label{tab:dems_comparison}
\end{table}

\begin{table}[t]
    \resizebox{\linewidth}{!}{%
    \begin{tabular}{lrrrr}
    \toprule
    {} & {} &  \textbf{Ext.} &  \textbf{No Ext.} &  \textbf{Diff.} \\
    \midrule
    \textbf{How frequently do} & Almost constantly   &       0.23 &          0.17 &        0.07 \\
    \textbf{you use Google?} & Several times a day &       0.42 &          0.35 &        0.07 \\
    & About once a day    &       0.11 &          0.12 &       -0.01 \\
    & A few times a week  &       0.13 &          0.15 &       -0.03 \\
    \midrule
    \textbf{How much of the} & All or almost all   &       0.06 &          0.05 &        0.01 \\
    \textbf{information you} & Most                &       0.33 &          0.19 &        0.14 \\
    \textbf{find using Google} & Some                &       0.39 &          0.36 &        0.03 \\
    \textbf{is accurate?}& Very little         &       0.11 &          0.19 &       -0.08 \\
    & None at all         &       0.02 &          0.08 &       -0.05 \\
    \midrule
    \textbf{Does Google} & Very accurate       &       0.35 &          0.27 &        0.08 \\
    \textbf{personalize} & Somewhat accurate   &       0.46 &          0.51 &       -0.05 \\
    \textbf{search results?} & Not very accurate   &       0.15 &          0.14 &        0.01 \\
    & Not at all accurate &       0.04 &          0.08 &       -0.04 \\
    \midrule
    \textbf{How satisfied you are} & Very satisfied      &       0.30 &          0.28 &        0.03 \\
    \textbf{with the search result} & Somewhat satisfied  &       0.62 &          0.60 &        0.02 \\
    \textbf{quality on Google?} & Not very satisfied  &       0.07 &          0.10 &       -0.03 \\
    \midrule
    \textbf{How much trust you} & A great deal        &       0.20 &          0.14 &        0.06 \\
    \textbf{have in information} & A moderate amount   &       0.70 &          0.69 &        0.00 \\
    \textbf{on Google?} & Not much            &       0.09 &          0.15 &       -0.06 \\
    \midrule
    \textbf{What do Google search} & Favor liberals      &       0.28 &          0.31 &       -0.03 \\
    \textbf{results favor:} & Favor conservatives &       0.06 &          0.06 &       -0.00 \\
    \textbf{liberals or conservatives?} & Neither             &       0.61 &          0.58 &        0.03 \\
    \midrule
    \textbf{Sample Size (weighted)} & {} & 790 & 1210 & {} \\
    \bottomrule
    \end{tabular}
    
    }
    \caption{Comparison of weighted Google usage and perceptions in the extension and no extension samples.}
    \label{tab:usage_comparison}
\end{table}

\begin{table}[t]
  \centering
  \footnotesize
  \begin{tabular}{lrr}
\toprule
\textbf{Search Engine} & \textbf{Subdomain} & \textbf{URL Param.} \\
\midrule
        google &               google.* &                        q \\
        google &          news.google.* &                        q \\
        google &        images.google.* &                        q \\
        google &        search.google.* &                        q \\
        google &          maps.google.* &                      -- \\
          bing &                 bing.* &                        q \\
         yahoo &                yahoo.com &                        p \\
         yahoo &           news.yahoo.com &                        p \\
         yahoo &         search.yahoo.com &                        p \\
         yahoo &   video.search.yahoo.com &                        p \\
         yahoo &      us.search.yahoo.com &                        p \\
         yahoo &    news.search.yahoo.com &                        p \\
         yahoo &       shopping.yahoo.com &                        p \\
         yahoo &  images.search.yahoo.com &                        p \\
    duckduckgo &           duckduckgo.com &                        q \\
        ecosia &               ecosia.org &                        q \\
          info &                 info.com &                        q \\
        yandex &               yandex.* &                     text \\
         baidu &                baidu.* &                       wd \\
         baidu &            map.baidu.* &                      -- \\
           ask &                  ask.com &                        q \\
           ask &        search.tb.ask.com &                        q \\
           aol &           search.aol.com &                        q \\
 searchencrypt &        searchencrypt.com &                        q \\
   privacywall &          privacywall.org &                        q \\
\bottomrule
\end{tabular} 
  \caption{Domains and search URL parameters for each horizontal search engine. The two map-related subdomains encoded queries in their URL path, not as a URL parameter. * indicates cases where we matched all \textbf{.com} and country-code top level domains.}
  \label{tab:hor_domains_and_url_parameter_key}
\end{table}

\subsection{Schema for Horizontal Search Engines}
\label{sec:horizontal-schema}

We manually developed search schema for the twelve major horizontal search engines that we identified using publicly available sources~\cite{ftc2013,cma2020,statcounter-search-usa,statista-search-usa,warren-2019-ballot}. We did so by visiting their homepages, running queries, and identifying the invariant portions of the URL. Additionally, many of the services that operate the twelve horizontal search engines also operate vertical search engines, \eg \path{news.google.com}, \path{maps.google.com}, and \path{mail.google.com} in the case of Google. We manually developed search schema for these vertical search engines using the same technique. Table~\ref{tab:hor_domains_and_url_parameter_key} shows the complete list of domains and their corresponding URL parameters for each search engine in our corpus.

\subsection{Assigning Searches to Vertical Segments}

\subsubsection{Google and Bing Products}
\label{sec:products}

We included searches on the following Google products in specific vertical segments:
\begin{enumerate}
    \item \path{youtube.com}: Streaming Media and Download
    \item \path{docs.google.com}, \path{photos.google.com},\\ \path{drive.google.com}: File Sharing and Storage
    \item \path{mail.google.com}: Web-based Email
    \item \path{google.com/search?tbm=nws}: News and Media\footnote{The \texttt{tbm} URL parameter redirects to a Google vertical search engine, \eg News or Shopping.}
    \item \path{google.com/search?tbm=shop}: Shopping
    \item \path{google.com/search?tbm=fin}: Finance and Banking
    \item \path{google.com/travel}: Travel
  
\end{enumerate}
We did not include searches on Google products (\eg Maps and Images) that did not fit cleanly into one of the top 30 vertical segments.

Similarly, we included searches on the following Bing products in specific vertical segments:
\begin{enumerate}
    \item \path{bing.com/news}: News and Media
    \item \path{bing.com/shop}: Shopping
    \item \path{bing.com/travel}, \path{bing.com/travelguide}: Travel
\end{enumerate}

We did not include searches on Bing products (\eg Maps and Images) that did not fit cleanly into one of the vertical segments. We could not include Outlook or Hotmail in the Web-based Email segment, or OneDrive in the File Sharing/Storage segment, because these services do not embed search queries in their URLs. 

\begin{table*}[t]
  \centering
  \footnotesize
  \begin{tabular}{lrl}
\toprule
                  \textbf{Segment} &  \textbf{Search Volume} &                       \textbf{Top Three Domains (Fraction of Search Volume)} \\
\midrule

                 Shopping & 113642 &  amazon.com (0.49);  google.com (0.22);  target.com (0.047) \\
           News and Media &  62112 &  bing.com (0.45);  google.com (0.45);   newspapers.com (0.073) \\
                 Business &  52067 &  google.com (0.45);  swagbucks.com (0.27);   bing.com (0.14) \\
                Reference &  46768 &  google.com (0.65);  bing.com (0.25);   findagrave.com (0.04)\\
        Social Networking &  40711 &  twitter.com (0.41);  facebook.com (0.29);  google.com (0.23) \\
Streaming Media and Download &  40389 &  youtube.com (0.49);  google.com (0.33);   bing.com (0.093) \\
          Web-based Email &  31891 &  mail.google.com (0.75);  mail.yahoo.com (0.24); google.com (0.0067) \\
                  Auction &  30456 &  ebay.com (0.94);  google.com (0.052);   bing.com (0.0052) \\
   Information Technology &  25736 &  google.com (0.74);  bing.com (0.21);  v10301.myubam.com (0.0076) \\
   
   Navigation	&  23236	&  google.com (0.83);  bing.com (0.17)  \\
                Education &  19335 &  google.com (0.79);  bing.com (0.14);  librarything.com (0.016) \\
            Entertainment &  18981 &  google.com (0.6);  bing.com (0.37);  imdb.com (0.015)\\
Government and Legal Organizations &  17839 &  google.com (0.55);  bing.com (0.24); forecast.weather.gov (0.16) \\
              Real Estate &  14490 &  zillow.com (0.78);  google.com (0.19);  bing.com (0.027) \\
              Pornography &  14452 &  xvideos.com (0.21);  xnxx.com (0.18);  google.com (0.16) \\
      Health and Wellness &  13280 &  google.com (0.78);  bing.com (0.16);  walgreens.com (0.028) \\
    Restaurant and Dining &  10460 &  google.com (0.6);  bing.com (0.26);  ubereats.com (0.02) \\
                    Games &  10357 &  google.com (0.4);  search.pch.com (0.36);  bing.com (0.092) \\
                   Travel &   9495 &  google.com (0.51);  bing.com (0.1);  kayak.com (0.067) \\
               Job Search &   9244 &  indeed.com (0.84);  google.com (0.097); bing.com (0.033) \\
      Finance and Banking &   8610 &  google.com (0.55);  finance.yahoo.com (0.29); bing.com (0.13) \\
                   Sports &   8241 &  google.com (0.6);  bing.com (0.4);  mlb.com (0.0013) \\
Newsgroups and Message Boards &   8085 &  google.com (0.51);  reddit.com (0.43);  bing.com (0.036) \\
   Society and Lifestyles &   5040 &  google.com (0.76);  bing.com (0.2);  craftsy.com (0.0066) \\
         Arts and Culture &   4134 &  google.com (0.52);  bing.com (0.14);  multcolib.overdrive.com (0.051) \\
 File Sharing and Storage &   3339&  photos.google.com (0.54);  drive.google.com (0.25);  unsplash.com (0.079) \\
        Personal Vehicles &   3227 &  google.com (0.72);  bing.com (0.15);  cargurus.com (0.1) \\
Personal Websites and Blogs &   3215 &  google.com (0.69);  bing.com (0.19);  tumblr.com (0.071) \\
    General Organizations &   3016 &  google.com (0.74);  bing.com (0.22);  opensecrets.org (0.022) \\
  Political Organizations &   1875  &  google.com (0.71);  bing.com (0.24);  politifact.com (0.019)\\

\bottomrule
\end{tabular}
  \caption{Top 30 vertical segments and the top three domains within each segment.}
  \label{tab:top_verticals_top_domains}
\end{table*}

\end{document}